\documentclass[11pt,a4paper]{article}
\usepackage{mlmodern}
\usepackage[T1]{fontenc}
\usepackage{verbatim}
\usepackage{float}
\usepackage{amsmath}
\usepackage{amssymb}
\usepackage{graphicx}
\usepackage{setspace}
\usepackage{subfig}
\usepackage[colorlinks=true,allcolors=indigo,backref]
 {hyperref}

\providecommand{\tabularnewline}{\\}

\usepackage{color}
\usepackage{setspace}
\usepackage{caption}
\newcommand{\mubar}{\bar{\mu}}
\usepackage{xcolor}
\usepackage{braket}
\usepackage{mathrsfs}
\usepackage{fullpage}
\usepackage{cite}
\usepackage{slashed}
\usepackage{epstopdf}
\usepackage[ddmmyyyy]{datetime}
\graphicspath{{./diagrams/}}

\allowdisplaybreaks
\definecolor{indigo}{rgb}{0.0, 0.25, 0.42}
\usepackage{cleveref}
\usepackage{array}

\renewcommand*{\backref}[1]{}
\renewcommand*{\backrefalt}[4]{({%
		\ifcase #1 Not cited.%
		\or Cited on page~#2%
		\else Cited on pages #2%
		\fi%
	})}

\newcommand{\E}{\epsilon}

\newcommand{\tbf}[1]{{\bf T}^{#1}}
\newcommand{\AsqNLP}[2]{	\left[\mathcal{A}^{#1}_{#2}\right]^2_{\text{NLP}}}
\newcommand{\AtsqNLP}[2]{	[\mathcal{\tilde{A}}^2]^{#1}_{#2}|_{\text{NLP}}}
\newcommand{\nn}{\nonumber \\ }

\newcommand{\xsctn}[2]	{\left.s_{12}^2\frac{d^2\sigma^{#1}_{#2}}{ds_{13}\,ds_{23}}\right|_{\text{NLP-LL}} }

\newcommand{\aq}{\bar{q}}
\newcommand{\amp}[2]{\mathcal{A}^{#1}_{#2}}
\newcommand{\anlp}[2]{\left.\mathcal{A}^{#1}_{#2}\right|_{\text{NLP}}}

\newcommand{\txb}[1]{\textcolor{indigo}{#1}}
\newcommand{\fac}{\left(\frac{\alpha_{s}}{6\pi v}\right)}
\newcommand{\angspinor}[1]{|#1\rangle}
\newcommand{\sqspinor}[1]{|#1]}

\newcommand{\cfq}{\left(\frac{\alpha_{s}}{6\pi v}\right)}
\newcommand{\slog}{\log\left( \frac{s_{45}}{\mu^2}\right )}
\newcommand{\alog}{\log\left( \frac{s_{12}s_{45}}{s_{13}s_{23}}\right )}

\global\long\def\Qqsa#1#2{\mathcal{Q}^{q_{#1}^{+}\aq_{#2}^{-}\to g_{#2}^{-}}}
\global\long\def\Qgqs#1#2{\mathcal{Q}^{g_{#1}^{+}q_{#2}^{+}\to q_{#1}^{+}}}

\global\long\def\Qqas#1#2{\mathcal{Q}^{q_{#1}^{+}\aq_{#2}^{-}\to g_{#1}^{+}}}
\global\long\def\Qasg#1#2{\mathcal{Q}^{\aq_{#1}^{-}g_{#2}^{-}\to \aq_{#2}^{-}}}

\global\long\def\alpnlp#1#2{\left.\mathcal{A}_{#2}^{#1}\right|_{\text{LP+NLP}}}%

\begin{document}
\begin{titlepage} 
\begin{flushleft}
Preprint 
\end{flushleft}

\begin{flushleft}
\vspace*{5cm}
 {\Large \bfseries Soft quark effects on H+jet production at NLP accuracy}
\bigskip
\bigskip
\medskip \\
\hrule height 0.05cm
\end{flushleft}
\begin{flushleft}
\vspace{1.0cm}
 \textbf{\large{\sffamily Sourav Pal and Satyajit Seth}} \medskip \\
\end{flushleft}

\begin{flushleft}
\textit{Theoretical Physics Division, Physical Research Laboratory,
}\\
\textit{Navrangpura, Ahmedabad 380009, India }
\end{flushleft}

\noindent \textit{E-mail: \href{mailto:sourav@prl.res.in}{sourav@prl.res.in},
\href{mailto:seth@prl.res.in}{seth@prl.res.in}
\medskip
}\\

\noindent \textsc{Abstract}: 
We define soft quark operators that enable construction of colour-ordered helicity amplitudes out of the non-radiative ones. We explore these operators for Higgs plus one jet production in a hadron collider and study the next-to-leading power corrections on all partonic channels involving quark(s) and/or anti-quark(s). We also investigate the effect of next-to-soft gluon radiation in these channels employing the technique illustrated  in~\cite{Pal:2023vec}. Analytical expressions of next-to-leading power leading logarithms thus obtained are simple and compact in nature. Further, we discuss the connection of such logarithms with that of the pseudo-scalar Higgs plus one jet production. 

\end{titlepage}

\hrule 
\tableofcontents \vspace{0.5cm}\hrule \vspace{0.4cm}

\section{Introduction}
Precise data obtained from the Large Hadron Collider, coupled with the absence of compelling new physics indications necessitate a better understanding of the Standard Model. Performing high precision perturbative Quantum Chromodynamics (QCD) calculations, either by incorporating higher order perturbative terms for fixed order corrections, or by performing resummation that account for specific enhanced logarithms across all orders in the perturbation series, are indispensable to understand the data obtained from the collider experiments. 

For all collider processes, one can define a threshold variable ($\xi$) that vanishes in the threshold limit and the differential cross-section may take the following form, 
\begin{align}
\frac{d\sigma}{d\xi}\,\approx\,\sum_{n=0}^{\infty}\alpha_{s}^{n}\left\{ \sum_{m=0}^{2n-1}C_{nm}\left(\frac{\log^{m}\xi}{\xi}\right)_{+}+d_{n}\delta(\xi)+\sum_{m=0}^{2n-1}D_{nm}\,\log^{m}\xi\right\} \,.\label{eq:gen-threshold}
\end{align}
The first set of logarithms with plus distributions and the delta functions appear due to the leading power (LP) approximation, whereas the second set of logarithms are the outcome of the next-to-leading power (NLP) approximation. 
The LP logarithms exhibit universality due to the factorisation of soft and collinear radiations from the hard interaction and as a consequence, resummation of these logarithms is well established in the literature. The pioneering works of refs.~\cite{Parisi:1980xd,Curci:1979am,Sterman:1987aj,Catani:1989ne,Catani:1990rp,Gatheral:1983cz,Frenkel:1984pz,Sterman:1981jc}
based on diagrammatics helped in devising methods for LP resummation. Subsequently several alternative methods of LP resummation were developed based on Wilson lines~\cite{Korchemsky:1993xv,Korchemsky:1993uz}, renormalisation
group (RG)~\cite{Forte:2002ni} and Soft Collinear Effective Theory
(SCET)~\cite{Becher:2006nr,Schwartz:2007ib,Bauer:2008dt,Chiu:2009mg}. A comprehensive comparison among different approaches can be
found in~refs.~\cite{Luisoni:2015xha,Becher:2014oda,Campbell:2017hsr}.

In view of the sizeable numerical impacts of NLP
logarithms~\cite{Kramer:1996iq,Ball:2013bra,Bonvini:2014qga,Anastasiou:2015ema,Anastasiou:2016cez,vanBeekveld:2019cks,vanBeekveld:2021hhv,Ajjath:2021lvg}, several methods have been proposed during the past decade to resum such logarithms~\cite{Grunberg:2009yi,Moch:2009hr,Moch:2009mu,Laenen:2010uz,Laenen:2008gt,deFlorian:2014vta,Presti:2014lqa,Bonocore:2015esa,Bonocore:2016awd,Laenen:2020nrt,DelDuca:2017twk,vanBeekveld:2019prq,Bonocore:2014wua,Bahjat-Abbas:2018hpv,Ebert:2018lzn,Boughezal:2018mvf,Boughezal:2019ggi,Bahjat-Abbas:2019fqa,Ajjath:2020ulr,Ajjath:2020sjk,Ajjath:2020lwb,Ahmed:2020caw,Ahmed:2020nci,Ajjath:2021lvg,Kolodrubetz:2016uim,Moult:2016fqy,Feige:2017zci,Beneke:2017ztn,Beneke:2018rbh,Bhattacharya:2018vph,Beneke:2019kgv,Bodwin:2021epw,Moult:2019mog,Beneke:2019oqx,Liu:2019oav,Liu:2020tzd,Boughezal:2016zws,Moult:2017rpl,Chang:2017atu,Moult:2018jjd,Beneke:2018gvs,Ebert:2018gsn,Beneke:2019mua,Moult:2019uhz,Liu:2020ydl,Liu:2020eqe,Wang:2019mym,Beneke:2020ibj,vanBeekveld:2021mxn}. Colour singlet production processes exhibit a universal pattern~\cite{DelDuca:2017twk}, however the universality of such terms is yet to be recognised for processes with coloured particles in the final state. To come up with a global resummation formulae, it is necessary to study jet-associated processes in order to understand the nature of NLP logarithms that originates from the emission of next-to-soft gluons and soft (anti-)quarks. Very few processes with a single coloured particle in the final state are studied so far~\cite{vanBeekveld:2019prq,Sterman:2022lki,Boughezal:2019ggi,vanBeekveld:2023gio} at the NLP accuracy and such scarcity of results arose due to the complexity in applying the method of momenta shifting at the squared amplitude level~\cite{vanBeekveld:2019prq}. Very recently, we have reported the NLP results of H+jet production via gluon fusion~\cite{Pal:2023vec} and shown that the shifting of dipole spinors in the non-radiative helicity amplitudes uniquely determines all colour-ordered radiative amplitudes at the NLP accuracy by capturing the effect of a next-to-soft gluon radiation. In fact, the method developed in~\cite{Pal:2023vec} is generic in nature and can be applied for processes with multiple jets in the final state.   

As discussed in~\cite{Pal:2023vec}, a blended mechanism comprising of helicity amplitudes and soft theorems of gauge theories can be used to obtain the NLP leading logarithms by shifting relevant spinors of the colour dipoles in non-radiative helicity amplitudes. However, as soft theorems are formulated specifically for the emission of massless bosons, they do not account for the generation of NLP logarithms resulting from the emission of soft (anti-)quarks. In this article, we study the NLP logarithms that emerge due to the soft quark radiation. We define soft quark operators that act on the colour-ordered non-radiative helicity amplitudes, thereby making the calculation simple and tractable. Using these operators, we look into the Higgs plus one jet production process and compute NLP leading logarithms arising from different partonic channels that involve a soft quark or anti-quark.  

The structure of our paper is as follows. In section~\ref{sec:sns},
we review the radiation of soft and next-to-soft gluon in terms of spinor shifts. Next, we construct operators for soft (anti-)quark radiation in section~\ref{sec:sqop}. In section~\ref{sec:nlpamp}, using these operators, we calculate NLP amplitudes for all the partonic channels that take part in Higgs plus one jet production and contain colour triplet particles in the initial and/or final states. Finally, we perform phase space integration over the squared NLP amplitudes to obtain the NLP leading logarithms for each of these channels in section~\ref{sec:nlplog}. Throughout this study, we have used a combination of in-house routines based on \texttt{QGRAF}~\cite{Nogueira:1991ex}, \texttt{FORM}~\cite{Vermaseren:2000nd} and \texttt{Mathematica}~\cite{Mathematica} to calculate all amplitudes and to perform phase space integrations. Feynman diagrams in this article are drawn using \texttt{FeynGame-2.1}~\cite{Harlander:2024qbn}.

\section{Next-to-soft gluon radiation}
\label{sec:sns}

In this section, we briefly review the emission of next-to-soft gluon in terms of spinor shifts in the non-radiative helicity amplitudes. 

The LP and NLP scattering amplitudes for $(n+1)$ particles with one next-to-soft gluon carrying momentum $p_s$ and helicity \lq $\displaystyle +$\rq \, can be expressed in terms $n$-particle scattering amplitude as~\cite{Luo:2014wea,Casali:2014xpa,Pal:2023vec},
\begin{align}
	\mathcal{A}_{\,n+1}^{\text{LP+NLP}}\bigg(\left\{ \lambda\angspinor s,\sqspinor s\right\} ,\left\{ \angspinor 1,\sqspinor 1\right\} ,\ldots,\left\{ \angspinor n,\sqspinor n\right\} \bigg) & \,=\,\frac{1}{\lambda^{2}}\frac{\braket{1n}}{\braket{1s}\braket{ns}}\times\mathcal{A}_{\,n}\bigg(\left\{ \angspinor 1,\sqspinor{1^{\prime}}\right\} ,\ldots,\left\{ \angspinor n,\sqspinor{n^{\prime}}\right\} \bigg)\,,\label{eq:LPplusNLP}
\end{align}
where 
\begin{equation}
	\sqspinor{1^{\prime}}\,=\,\sqspinor 1+\Delta_{s}^{(1,n)}\sqspinor s\,,\nn\sqspinor{n^{\prime}}\,=\,\sqspinor n+\Delta_{s}^{(n,1)}\sqspinor s\,,\label{eq:gen-shifts}
\end{equation}
and, 
\begin{equation}
	\Delta_{s}^{(i,j)}=\lambda\frac{\braket{js}}{\braket{ji}}\,.\label{eq:mom-ratio}
\end{equation}
Here $\angspinor i$ and $\sqspinor i$ denote the holomorphic and
anti-holomorphic spinors associated with the particle $i$ carrying
momentum $p_{i}$. $\lambda$ is the scaling parameter by which the radiative gluon momentum is being scaled and the exponent of this parameter determines the leading and sub-leading terms that are of $\mathcal{O}(\lambda^{-2})$ and $\mathcal{O}(\lambda^{-1})$ respectively. To obtain the above formulae, we have used the holomorphic soft limit~\cite{Casali:2014xpa,Luo:2014wea} {\em i.e.,} 
\begin{align}
\angspinor s\rightarrow\lambda\,\angspinor s\,,\quad\sqspinor s\rightarrow\sqspinor s\,,
\end{align}
and the BCFW deformation of $s$ and $n$ pair when particle $1$ forms a three particle amplitude involving particle $s$ and the on-shell cut propagator.  
Note that, in the colour-ordered amplitude of eq.~\eqref{eq:LPplusNLP}, the soft gluon $s$ is placed in between particles $1$ and $n$, and that can essentially be calculated by shifting the anti-holomorphic spinors ({\em i.e.}, $\sqspinor 1$ and $\sqspinor n$) of the colour dipole $\mathcal{D}_{1n}$. In fact, for a $n$-gluon amplitude one can form $^nC_2=n(n-1)/2$ colour dipoles and in the next-to-soft limit, shifting of corresponding anti-holomorphic spinors in each dipole produces one independent colour-ordered amplitude that participates in the $(n+1)$-gluon scattering. When the helicity of the next-to-soft gluon is \lq$\displaystyle -$\rq \, one needs to replace all angle spinors with square spinors and vice versa in the above set of formulae.

\section{Soft quark radiation}
\label{sec:sqop} 
\begin{figure}[t]
\centering \subfloat[]{\includegraphics[scale=0.35]{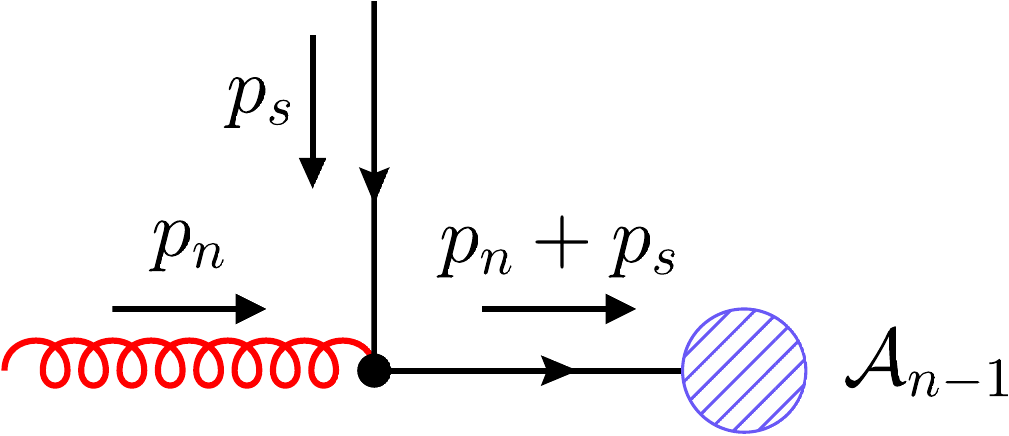} 
}\qquad\subfloat[]{\includegraphics[scale=0.35]{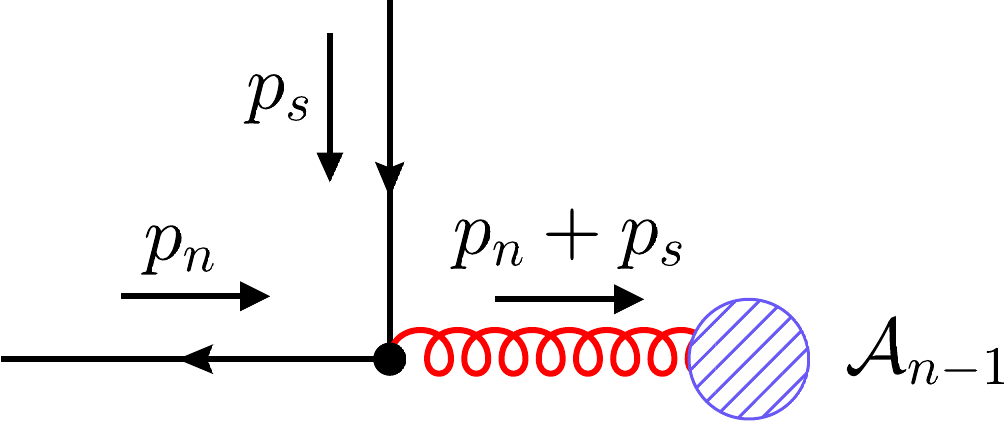} 
}
\caption{Feynman diagrams showing the radiation of soft quarks. Diagram (a)
shows
merging of a soft quark and a gluon; diagram (b) shows merging of a soft quark with an antiquark. }
\label{fig:sqfeynman} 
\end{figure}

We now turn our attention towards the emission of a soft quark or anti-quark from the non-radiative amplitudes. Our aim here is to express the colour-ordered $(n+1)$-particle amplitudes as a blend of soft quark operators and $n$-particle amplitudes; akin to eq.~\eqref{eq:LPplusNLP}.    
In order to construct soft quark operators, we consider combining of a soft quark to the adjacent hard colour particles, thereby forming new hard candidates to engage in non-radiative amplitudes. Such merging can occur in two possible ways: ($i$) a soft quark can get merged to an adjacent gluon and form a quark, as shown in fig.~(\ref{fig:sqfeynman}\txb{a}), ($ii$) a soft quark can get merged to an adjacent anti-quark and form a gluon, as depicted in  fig.~(\ref{fig:sqfeynman}\txb{b}). 

We first discuss the scenario shown in fig.~(\ref{fig:sqfeynman}\txb{a}) where a soft quark operator is to be formed by melding a soft quark with gluon. Note that we work on the colour ordered amplitude and consider one merging at a time which always involves one power of strong coupling. Therefore, for simplicity, the colour and coupling strength are stripped off the amplitude and then it is expressed as, 
\begin{align}
\amp{q_s g_n \to q}{n+1}\,&=\,\bar{u}_{h_{s}}(p_s)\gamma^{\mu}\frac{i(\slashed{p_{n}}+\slashed{p}_{s})}{(p_{n}+p_s)^{2}}\frac{-i}{\sqrt{2}}\mathcal{E}^{\mu}(p_{n},h_{n})\,\amp{}{}\,\nonumber \\
&=\,\frac{1}{\sqrt{2}}\bar{u}_{h_{s}}(p_s)\slashed{\mathcal{E}}(p_{n},h_{n})\frac{\slashed{p}_{n}}{2p_{n}\cdot p_{s}}\mathcal{A}\,.\label{eq:sqfeyn} 
\end{align}
where the $n$-particle Born amplitude in the soft quark limit can be written as, 
\begin{align}
\amp{}{n}\,=\,\bar{u}_{h_{n}}(p_{n})\, \amp{}{}\,.
\end{align}
All the remaining $(n-1)$ particles are kept implicit throughout as they do not intervene in the present calculation directly. We recall that the dimension of $\bar{u}_{h_{s}}(p_s)$ is $\mathcal{O}(\sqrt{p_s})$ and that leaves the dimension of the overall amplitude presented in eq.~\eqref{eq:sqfeyn} as $\mathcal{O}(1/{\sqrt{p_s}})$ in the soft quark limit. Therefore, the squared amplitude, although it does not generate any LP threshold contribution, can produce NLP logarithms. 

We replace $\slashed{p}_n$ in eq.~\eqref{eq:sqfeyn} as $\slashed{p}_n=\sum\limits_{h_n}u_{h_n}(p_n)\bar{u}_{h_n}(p_n)$ and then it is straightforward to show that, 
\begin{align}
\amp{q_s^+ g_n^+\to q^+}{n+1}\,=\,-\frac{1}{\braket{ns}}\,\mathcal{A}_{n}\,,\qquad
\amp{q_s^- g_n^-\to q^-}{n+1}\,=\,\frac{1}{[ns]}\,\mathcal{A}_{n}\,,
\label{eq:sqfeynman2}
\end{align}
and QCD being a helicity conserving theory, all the remaining combinations that can arise due to other helicities of the mingling soft quark and gluon become zero. 

A similar calculation can be performed for fig.~(\ref{fig:sqfeynman}\txb{b}),
where a soft quark merges with an anti-quark and produces a hard gluon to enter in $\mathcal{A}_n$. In this case, the non-zero contributions can be represented as, 
\begin{align}
\amp{q_s^+ \bar{q}_n^-\to g^-}{n+1}\,=\,\frac{1}{\braket{ns}}\,\mathcal{A}_{n}\,,\qquad
\amp{q_s^- \bar{q}_n^+\to g^+}{n+1}\,=\,-\frac{1}{[ns]}\,\mathcal{A}_{n}\,.
\label{eq:sqfeynman3}
\end{align}
The effects of radiation of soft anti-quarks can easily be obtained in a similar fashion as above. 

We slightly alter the notation to be more explicit and represent such factorisations in terms soft (anti-)quark operators ($\mathcal{Q}$) as follows,
\begin{align}
\amp{s^{i} h^{j}}{n+1}\,=\,\mathcal{Q}^{s^{i} h^{j}\to c^{k}}\,\amp{c^{k}}{n}\,,\qquad
\label{eq:sqfeynman4}
\end{align}
where $s^i$ denotes the ``{\em soft}\,'' (anti-)quark with helicity $i$; $h^j$ is the ``{\em hard}\,'' particle with helicity $j$ and the ``{\em clubbed}\,'' colour particle $c^k$ with helicity $k$ is being formed by combining the hard and soft particles of the $(n+1)$-particle amplitude. Note that the clubbed particle carries the helicity and momentum of the hard particle, but not its colour and therefore $j=k$ always. 
Two independent non-zero operators are listed in Table~\eqref{tab:sqtable} and all the rest of the non-zero operators, including the soft anti-quark operators, are charge and/or parity conjugate of these two. Except $q\bar{q}$ merging, interchanging of the position of the soft and hard particles may incur an additional negative sign in the soft (anti-)quark operators to take care of a definite colour ordering. 

\begin{table}[tbh]
\begin{center}
	\begin{onehalfspacing}
\begin{tabular}[t]{|c|c|c|}
\hline 
Soft Quark Operator & Explicit Form\tabularnewline
\hline
$\mathcal{Q}^{q_{s}^{+}\aq_{h}^{-}\to g_{c}^{-}}$ & ${\displaystyle -\frac{1}{\braket{sh}}}$\tabularnewline
\hline 
$\mathcal{Q}^{g_{h}^{+} q_{s}^{+}\to q_{c}^{+}}$ & ${\displaystyle -\frac{1}{\braket{sh}}}$\tabularnewline
\hline 
\end{tabular}\caption{Two independent soft quark operators for two different types of merging. All other non-zero combinations including the soft anti-quark operators can be obtained by taking appropriate parity and/or charge conjugations. \label{tab:sqtable} }
	\end{onehalfspacing}
\end{center}
\end{table}

Note that the soft quark operators were first defined in~\cite{vanBeekveld:2019prq}, where the NLP amplitudes were calculated using conventional Feynman diagrammatic approach and they were also used in~\cite{vanBeekveld:2023liw} to explain exponentiation of the soft quark effects. Those operators exhibit intricate colour structures, thereby rendering them challenging to employ in practical computations. However, colour ordered helicity amplitudes together with the soft quark operators discussed above remarkably streamline the computation of NLP amplitudes and hence the derivation of NLP logarithms. We apply this technique to acquire NLP logarithms that originate due to soft (anti-)quark radiations in the case of Higgs plus one jet production at the LHC.

\section{H+jet production}
\label{sec:nlpamp}
The most dominant production mechanism of Higgs boson at the LHC is through gluon
fusion. Though gluons do not interact directly with the Higgs boson, it can be produced via a massive quark loop. Since top quark
is the heaviest among quarks, the dominant contribution comes from the top quark loop. In the limit $m_{t}\rightarrow\infty$, the top
quark effect can be integrated out to obtain an effective Lagrangian
\cite{Wilczek:1977zn, Shifman:1978zn} given by, 
\begin{align}
\mathcal{L}_{\,\text{eff}}\,=\,-\frac{1}{4}\,G\,\,H\,\text{Tr}(F_{\mu\nu}^{a}F^{\mu\nu,a})\,,
\end{align}
where $F_{\mu\nu}^{a}$ is the QCD field strength tensor and the effective
coupling at the lowest order is given by $G=\alpha_{s}/3\pi v$. Here $\alpha_{s}$
denotes the strong coupling constant and $v$ is the vacuum expectation value of the Higgs field. 

We stick to this effective Lagrangian and study the soft (anti-)quark and the next-to-soft gluon radiation effects on Higgs plus one
hard jet production at the LHC. Therefore, the process that we are interested in involves Higgs plus four partons among which one is either a soft (anti-)quark or a next-to-soft gluon. NLP amplitudes that arise from Higgs plus 4-gluon channel are thoroughly discussed in~\cite{Pal:2023vec}. There are two other types of channels that involve Higgs and four partons: $(i)$  one quark anti-quark pair with two gluons and Higgs $\left(q\aq ggH\right)$, and $(ii)$
two pairs of quark anti-quark and a Higgs $\left(q\aq Q\bar{Q}H\right)$. Scattering amplitudes of these processes are given in~\cite{Kauffman:1996ix}. 

NLP contributions of Higgs boson plus one jet production can come from four 
different scattering processes that fall into the first category {\em i.e.,} $q\aq ggH$ and those four processes are given by, 
\begin{align}
q(p_{1})+\aq(p_{2})+H(-p_{3})+g(-p_{4})+g(-p_{5})\rightarrow0 & \,\,,\label{eq:qqHgg}\\
q(p_{1})+g(p_{2})+H(-p_{3})+\aq(-p_{4})+g(-p_{5})\rightarrow0 & \,,\label{eq:qqHgg1}\\
\aq(p_{1})+g(p_{2})+H(-p_{3})+q(-p_{4})+g(-p_{5})\rightarrow0 & \,,\label{eq:qqHgg2}\\
g(p_{1})+g(p_{2})+H(-p_{3})+q(-p_{4})+\aq(-p_{5})\rightarrow0 & \,. \label{eq:qqHgg3}
\end{align}
NLP logarithms can appear when either one of the partons with momentum $p_4$ and $p_5$ becomes (next-to-)soft. However, at the scattering amplitude level, all these four processes are related to one another by simple relabelling of their momenta. For simplicity, one can always consider all external particles as incoming and calculate NLP amplitudes for any one out of these four processes. 

The second kind of scattering process that contributes to NLP logarithms involves a Higgs boson and two pairs of quark and anti-quark,    
\begin{align}
q(p_{1})+\aq(p_{2})+H(-p_{3})+Q(-p_{4})+\bar{Q}(-p_{5})\rightarrow0 & \,.
\label{eq:H4q}
\end{align}
Here $q$ and $Q$ denote different quark flavours. This process has only one independent helicity amplitude which is given in~\cite{Kauffman:1996ix} and when all (anti-)quarks are of same flavour, one needs to add an additional contribution by replacing $p_2\leftrightarrow p_5$. 

\subsection{Colour-ordered amplitudes}
\label{sec:corder}
Scattering  amplitude of $n$-gluons with one Higgs boson can be expressed as, 
\begin{equation}
	\mathcal{A}_{nH} (\{p_i,h_i,c_i\})\,=\, -i\,\left(\frac{\alpha_{s}}{6\pi v}\right)g_s^{n-2} \sum_{\sigma \in \mathcal{S}_{n'}} \text{Tr} \left( \tbf{c_1}\tbf{c_2}\tbf{c_4} \ldots  \tbf{c_{n+1}} \right )\mathcal{A}^{h_1\,h_2\,h_4\,\ldots\,h_{n+1}}_{1_g\,2_g\,H\,4_g\, \ldots \,(n+1)_g} \, ,
	\label{eq:gen-ng}
\end{equation}
where $ \mathcal{S}_{n'} $ denotes $ (n-1)! $ non-cycling permutations of $n$-gluons. $\tbf{c_i}$ represents SU(3) colour matrix in the fundamental representation and such generators are normalized as $ \text{Tr} (\tbf{c_1}\tbf{c_2})= \delta^{c_1 c_2} $. We can write the full amplitude of 4-gluons and Higgs for a given helicity configuration in terms of colour-ordered amplitudes as, 
\begin{align}
	& \mathcal{A}_{4H}(\{p_i,h_i,c_i\})\,\nn& =\,-i \, \left (\frac{\alpha_{s}}{6\pi v}\right )\, g_s^2 \Bigg[ \left \{\text{Tr} \left (\tbf{c_1}\tbf{c_2}\tbf{c_4}\tbf{c_5}\right )+ \text{Tr} \left (\tbf{c_1}\tbf{c_5}\tbf{c_4}\tbf{c_2}\right )\right \} \mathcal{A}^{h_1\,h_2\,h_4\,h_5}_{1_g\,2_g\,4_g\,5_g} \nn & 
	+\left \{\text{Tr} \left (\tbf{c_1}\tbf{c_4}\tbf{c_5}\tbf{c_2}\right )+ \text{Tr} \left (\tbf{c_1}\tbf{c_2}\tbf{c_5}\tbf{c_4}\right )\right \} \mathcal{A}^{h_1\,h_2\,h_4\,h_5}_{1_g\,4_g\,5_g\,2_g\,} \nn &
	+\left \{\text{Tr} \left (\tbf{c_1}\tbf{c_5}\tbf{c_2}\tbf{c_4}\right )+ \text{Tr} \left (\tbf{c_1}\tbf{c_4}\tbf{c_2}\tbf{c_5}\right )\right \} \mathcal{A}^{h_1\,h_2\,h_4\,h_5}_{1_g\,5_g\,2_g\,4_g} \Bigg] \,.
	\label{eq:gen4gamp}
\end{align}
For brevity, we avoid writing H explicitly in the colour-ordered amplitudes here and in the rest of the paper. Squaring the above equation and summing over colours we obtain, 
\begin{align}
	\sum_{\text{colours}}|\mathcal{A}_{4H}(\{p_i,h_i,c_i\})|^2&=\bigg[\left (\frac{\alpha_{s}}{6\pi v}\right ) g_s^2\bigg]^2 (N^2-1) \bigg \{ 2 N^2 \left ( \left|\mathcal{A}^{h_1\,h_2\,h_4\,h_5}_{1_g\,2_g\,4_g\,5_g}\right|^2+\left|\mathcal{A}^{h_1\,h_2\,h_4\,h_5}_{1_g\,4_g\,5_g\,2_g}\right|^2+\left|\mathcal{A}^{h_1\,h_2\,h_4\,h_5}_{1_g\,5_g\,2_g\,4_g}\right|^2\right) \nn
	&  -4 \frac{(N^2-3)}{N^2} \left|\mathcal{A}^{h_1\,h_2\,h_4\,h_5}_{1_g\,2_g\,4_g\,5_g}+\mathcal{A}^{h_1\,h_2\,h_4\,h_5}_{1_g\,4_g\,5_g\,2_g}+\mathcal{A}^{h_1\,h_2\,h_4\,h_5}_{1_g\,5_g\,2_g\,4_g}\right|^2 \bigg\}\,.
	\label{eq:genAsq}
\end{align}
in which the sub-leading colour term given in the second line of the above equation does not contribute at all due to the dual Ward identity~\cite{Dixon:2004za,Kauffman:1996ix}.

The general form of scattering amplitude involving a quark antiquark pair, $n$ gluons and a Higgs boson is given by, 
\begin{align}
\amp{q\bar{q}}{nH}(\{p_i,h_i,c_i\})\,=\,-i\fac g_s^{n}\sum_{\sigma\in S_{n}}(\tbf{c_{4}}\ldots\tbf{c_{n+3}})_{j_{2}j_1}{}\amp{h_1\,h_2\,h_4\,\ldots\,h_{n+3}}{1_{q}\,2_{\aq}\,H\,4_g\,\ldots\,(n+3)_g}\,,\label{eq:genqaqgh}
\end{align}
where $S_{n}$ is the $n!$ possible non-cyclic permutations of $n$ gluons in the presence of a $q\bar{q}$ pair. 

Now, using the above equation, the amplitude for the process given in eq.~\eqref{eq:qqHgg} can be expressed as, 
\begin{align}
	\mathcal{A}^{q\bar{q}}_{2H}(\{p_i,h_i,c_i\})\,=\,-i\fac g_{s}^{2}\left[\left(\tbf{c_{4}}\tbf{c_{5}}\right)_{j_{2}j_{1}}\amp{h_1\,h_2\,h_4\,h_5}{1_{q}\,2_{\aq}\,4_g\,5_{g}}+\left(\tbf{c_{5}}\tbf{c_{4}}\right)_{j_{2}j_{1}}\amp{h_1\,h_2\,h_4\,h_5}{1_{q}\,2_{\aq}\,5_g\,4_g}\right]\,,
\end{align}
and squaring the above amplitude and summing over colours we obtain, 
\begin{align}
\sum_{\text{colours}}	\left|\mathcal{A}^{q\bar{q}}_{2H}(\{p_i,h_i,c_i\})\right|^2\,&=\,\fac^{2}g_{s}^{4}\,(N^{2}-1)\bigg\{N\left(\left|\amp{h_1\,h_2\,h_4\,h_5}{1_{q}\,2_{\aq}\,4_g\,5_{g}}\right|^{2}+\left|\amp{h_1\,h_2\,h_4\,h_5}{1_{q}\,2_{\aq}\,5_g\,4_{g}}\right|^{2}\right) \nn
&-\frac{1}{N}\left|\amp{h_1\,h_2\,h_4\,h_5}{1_{q}\,2_{\aq}\,4_g\,5_g}+\amp{h_1\,h_2\,h_4\,h_5}{1_{q}\,2_{\aq}\,5_g\,4_g}\right|^{2}\bigg\}\,.
\label{eq:qaqHggsq}
\end{align}

The colour-ordered amplitude for Higgs  plus two quarks and two anti-quarks is given by,
\begin{align}
	\amp{q\aq Q\bar{Q}}{H}(\{p_i,h_i,c_i\})\,=\, -i \cfq \, g_s^2\, (\tbf{c_1})_{j_2 j_1} (\tbf{c_1})_{j_4 j_3}\,  \amp{h_1\,h_2\,h_4\,h_5}{1_q\,2_{\aq}\,4_Q\,5_{\bar{Q}}}
\label{eq:H4q}
\end{align}
where the helicities of the antiquarks are fixed by the helicities of the quarks {\em i.e.,} $h_2=-h_1$ and $h_4=-h_5$. Squaring the amplitude and summing over colours we get, 
\begin{align}
\sum_{\text{colours}} |\amp{q\aq Q\bar{Q}}{H}(\{p_i,h_i,c_i\})|^2\,=\, \cfq^2 g_s^4\, (N^2-1) \left|\amp{h_1\,h_2\,h_4\,h_5}{1_q\,2_{\aq}\,4_Q\,5_{\bar{Q}}} \right|^2\,,
\end{align} 
and here different two different quark flavours are considered. For the case where quarks are of same flavours, we can write the squared amplitude as,  
\begin{align}
	\sum_{\text{colours}} |\amp{q\aq Q\bar{Q}}{H}(\{p_i,h_i,c_i\})|^2\,&=\, \cfq^2 g_s^4\, (N^2-1)  \bigg\{ \left|\amp{h_1\,h_2\,h_4\,h_5}{1_q\,2_{\aq}\,4_Q\,5_{\bar{Q}}} \right|^2+\left|\amp{h_1\,h_2\,h_4\,h_5}{1_q\,2_{\aq}\,4_q\,5_{\bar{q}}}\right|^2 \nn
	&+\frac{\delta_{h_1h_4}}{N}  \bigg(\amp{h_1\,h_2\,h_4\,h_5}{1_q\,2_{\aq}\,4_Q\,5_{\bar{Q}}}\amp{\dagger\,h_1\,h_2\,h_4\,h_5}{1_q\,2_{\aq}\,4_q\,5_{\bar{q}}}+\amp{h_1\,h_2\,h_4\,h_5}{1_q\,2_{\aq}\,4_q\,5_{\bar{q}}}\amp{\dagger\,h_1\,h_2\,h_4\,h_5}{1_q\,2_{\aq}\,4_Q\,5_{\bar{Q}}} \bigg) 
	\bigg\}\,,
\label{eq:flavoursq}	
\end{align} 
where $\amp{h_1\,h_2\,h_4\,h_5}{1_q\,2_{\aq}\,4_q\,5_{\bar{q}}}=\amp{h_1\,h_2\,h_4\,h_5}{1_q\,5_{\aq}\,4_Q\,2_{\bar{Q}}}$, and the sub-leading colour term exists only when the quarks carry the same helicity. 
\subsection{Leading order amplitudes}
The next-to-soft gluon effects can be obtained by shifting appropriate spinors in the leading order amplitudes. On the other hand, the soft quark effects can be obtained through soft quark operators which eventually result into spinor factors multiplied with the leading order amplitudes. So the basic ingredients to compute NLP amplitudes are the leading order helicity amplitudes that contain Higgs plus three partons and we calculate them here under.  

The leading order process with a quark anti-quark pair, a Higgs, and a gluon is given by,  
\begin{align}
q(p_{1})+\aq(p_{2})+H(p_{3})+g(p_{4})\rightarrow0\,.
\label{eq:qaqgHkin}
\end{align}
Note that, for simplicity, we consider all external legs as incoming throughout this article unless mentioned otherwise. There is only one independent colour-ordered helicity amplitude for this process, and that is 
given by, 
\begin{align}
\amp{+-+}{1_{q}\,2_{\aq}\,4_g}\,=\,\frac{\,\,[14]^{2}}{[12]}\,.\label{eq:qaqglo}
\end{align}
The remaining helicity amplitudes for this process can be obtained by
taking either parity or charge conjugation of the above amplitude.

The other leading order process which involves Higgs boson and three gluons can be represented as,   
\begin{align}
g(p_{1})+g(p_{2})+H(p_{3})+g(p_{4})\rightarrow0\,,
\label{eq:gggHkin}
\end{align}
and two independent colour-ordered helicity amplitudes for this process are, 
\begin{align}
\mathcal{A}_{1_g\,2_g\,4_g}^{+++} & =\frac{m_{H}^{4}}{\braket{12}\braket{24}\braket{41}}\,,\qquad\mathcal{A}_{1_g\,2_g\,4_g}^{-++}=\frac{[24]^{3}}{[12][14]}\,.\label{eq:ggglo}
\end{align}
All other helicity combinations for this process can be constructed out of these two helicity configurations. All these amplitudes agree with the results given in~\cite{Kauffman:1996ix}. 

\subsection{NLP amplitudes for $ q\aq ggH $}
We present here the NLP amplitudes for the scattering involving Higgs, a quark antiquark pair, and two gluons. Throughout this section, we choose the kinematic configuration given in eq.~\eqref{eq:qqHgg}, except the fact that all external particles are now incoming. 
Due to the fact that quark and anti-quark always possess opposite helicities, there are total $ \displaystyle 2^3=8 $ helicity configurations. Out of these eight, one needs to calculate only three configurations, since the remaining configurations can be obtained by using parity and charge conjugation. We choose to calculate NLP amplitudes for the following three independent helicity configurations: ($i$) $\amp{+-++}{1_{q}\,2_{\aq}\,4_g\,5_g}$, ($ii$) $\amp{+-+-}{1_{q}\,2_{\aq}\,4_g\,5_g}$, ($iii$) $\amp{+--+}{1_{q}\,2_{\aq}\,4_g\,5_g}$, and their full results are given by~\cite{Kauffman:1996ix}, 
\begin{align}
\amp{+-++}{1_{q}\,2_{\aq}\,4_g\,5_g} &= 
\frac{\langle 2|1+5|4]^2}{s_{125}} \frac{[15]}{\langle 25 \rangle}\left( \frac{1}{s_{12}} + \frac{1}{s_{15}} \right)
-\frac{\langle 2|1+4|5]^2}{s_{124}\, s_{12}} \frac{[14]}{\langle 24 \rangle}
+\frac{\langle 2|4+5|1]^2}{[12] \langle 24 \rangle \langle 25 \rangle \langle 45 \rangle} 
\,,\nn
\amp{+-+-}{1_{q}\,2_{\aq}\,4_g\,5_g} &= 
\frac{\langle 25 \rangle^3}{\langle 12 \rangle \langle 24 \rangle \langle 45 \rangle}
-\frac{[14]^3}{[12] [15] [45]}  \,,\nn
\amp{+--+}{1_{q}\,2_{\aq}\,4_g\,5_g} &= 
\frac{\langle 24 \rangle^2 \langle 14 \rangle}{\langle 12 \rangle \langle 15 \rangle \langle 45 \rangle}
-\frac{[15]^2 [25]}{[12] [24] [45]}  \,,
\label{eq:fullqaHgg}
\end{align}
where $s_{ijk}=(s_{ij}+s_{jk}+s_{ik})$ and $s_{ij}=2 p_i.p_j$, when $p_i$ and $p_j$ denote momenta of massless particles.

\subsubsection{Radiation of soft quark}
\label{sec:sq}
Soft quark contribution can be obtained by expanding the $q\aq ggH$ helicity amplitudes given in eq.~\eqref{eq:fullqaHgg} in powers of the holomorphic quark spinor and then keeping the leading terms only. Alternatively, they can also be obtained using the factorisation of amplitudes in terms of soft quark operators, defined in section~\ref{sec:sqop}. In what follows, we take the latter approach to express the NLP amplitudes due to soft quark radiation. In every case, these results perfectly match with the ones derived using the former approach. 

We choose to calculate NLP amplitudes for the process given in eq.~\eqref{eq:qqHgg}, provided all the external particles are now incoming. As discussed in section~\ref{sec:sqop}, in a colour-ordered helicity amplitude the soft quark can couple either to an adjacent anti-quark or to an adjacent gluon, and can produce a {\em clubbed} gluon or quark. Following that, the independent NLP amplitudes can be expressed as, 
\begin{align}
\anlp{+-++}{1_{q(s)}\,2_{\aq}\,4_g\,5_g}\,=\, & \Qqsa{1}{2}\amp{-++}{2_g\,4_g\,5_g}+\Qgqs{5}{1}\amp{+-+}{5_{q}\,2_{\aq}\,4_g}\,,\nn
\anlp{+-+-}{1_{q(s)}\,2_{\aq}\,4_g\,5_g}\,=\, & \Qqsa{1}{2}\amp{-+-}{2_g\,4_g\,5_g}\,,\nn
\anlp{+--+}{1_{q(s)}\,2_{\aq}\,4_g\,5_g}\,=\, & \Qqsa{1}{2}\amp{--+}{2_g\,4_g\,5_g}+\Qgqs{5}{1}\amp{+--}{5_{q}\,2_{\aq}\,4_g}\,.
\label{eq:sqqaqHgg}
\end{align}
Explicit forms of the soft quark operators that take part in the above equations are provided in Table~\ref{tab:sqtable}. Note that the second NLP amplitude contains only one single operator as the operator of type $\mathcal{Q}^{g_{h}^{-} q_{s}^{+}\to q_{c}^{-}}=0$.

\subsubsection{Radiation of soft anti-quark}
\label{sec:saq}
A soft anti-quark can either couple to a hard quark and form a {\em clubbed} gluon, or it can be combined to a hard  gluon to produce a {\em clubbed} anti-quark. 
The three different helicity amplitudes in case of soft anti-quark radiation can be written as,
\begin{align}
\anlp{+-++}{1_{q}\,2_{\aq(s)}\,4_g\,5_g}\,=\, & \Qqas{1}{2}\amp{+++}{1_g\,4_g\,5_g}\,,\nn
\anlp{+-+-}{1_{q}\,2_{\aq(s)}\,4_g\,5_g}\,=\, & \Qqas{1}{2}\amp{++-}{1_g\,4_g\,5_g}\,,\nn
\anlp{+--+}{1_{q}\,2_{\aq(s)}\,4_g\,5_g}\,=\, & \Qqas{1}{2}\amp{+-+}{1_g\,4_g\,5_g}+\Qasg{2}{4}\amp{+--}{1_{q}\,4_{\aq}\,5_g}\,.
\label{eq:sqqaqHgg}
\end{align}
The generic forms of the two types of non-zero soft anti-quark operators that play the part here are, $\mathcal{Q}^{q_{h}^{+} \aq_{s}^{-}\to g_{c}^{+}}=\frac{1}{[sh]}$ and $\mathcal{Q}^{\aq_{s}^{-} g_{h}^{-}\to \aq_{c}^{-}}=-\frac{1}{[sh]}$, and the same can be inferred once appropriate conjugations on the ingredients given in Table~\ref{tab:sqtable} are applied. Again, we find perfect agreement when the same set of results are produced by retaining only the leading terms in the expansion of eq.~\eqref{eq:fullqaHgg} in powers of the anti-holomorphic anti-quark spinor.

\subsubsection{Radiation of soft gluons}
\label{sec:nsg}
The colour-ordered amplitudes for soft gluon radiation can be obtained
very easily through formation of colour dipoles and shifting
of appropriate spinors in the leading order amplitudes\cite{Pal:2023vec}. The helicity of the emitted soft gluon can be both \lq$\displaystyle+$\rq \, or \lq$\displaystyle-$\rq \, and therefore configurations that are maximally helicity violating (MHV) may also appear. However, it is to be noted that any MHV configuration does not contribute to the NLP amplitude. We describe this by considering a simple example. Let us consider that a gluon with
\lq$\displaystyle-$\rq \, helicity is being emitted from $\amp{+-+}{1_{q}\,2_{\aq}\,4_g}$. Now following the arguments illustrated in section~\ref{sec:sns}, one needs to shift the holomorphic spinors depending on where the soft gluon is going to be attached. However, the leading order amplitude given in eq.~\eqref{eq:qaqglo} does not contain any holomorphic spinor at all and hence NLP amplitudes of $\amp{}{1_{q}^{+}2_{\aq}^{-};4_g^+5_g^-}$ vanishes. The same can be verified by expanding the full result given in eq.~\eqref{eq:fullqaHgg} in terms of anti-holomorphic spinor $|5]$. We note in passing that no NMHV amplitudes contribute at the NLP the case of Higgs plus one jet production via gluon fusion~\cite{Pal:2023vec}. Therefore vanishing of NLP amplitudes with alternative helicity signatures, is independent of the production channel when next-to-soft gluon radiation is considered. 

In the absence of MHV amplitudes at the NLP, we are left with only one type of non-vanishing amplitude $\amp{+-++}{1_{q}\,2_{\aq}\,4_g\,5_g}$. Its  various colour-ordered components, as given in eq.~\eqref{eq:qaqHggsq}, can be obtained by shifting appropriate spinors of the colour dipoles formed at the leading order. The LP and NLP contributions can be expressed as~\cite{Pal:2023vec}, 
\begin{align}
\alpnlp{+-++}{1_{q}\,2_{\aq}\,4_g\,5_g}\,=\,\frac{\braket{14}}{\braket{15}\braket{45}}\amp{+-+}{1^{'}_{q}\,2_{\aq}\,4^{'}_g}\,,
\label{eq:D14}
\end{align}
where the primes indicate the shifted spinors
$|1^{'}]$ and $|4^{'}]$ according to eq.~\eqref{eq:gen-shifts}. Similarly, the other
colour-ordered amplitude can be obtained by shifting the spinors $|2^{'}]$
and $|4^{'}]$, and that is given by, 
\begin{align}
\alpnlp{+-++}{1_{q}\,2_{\aq}\,5_g\,4_g}\,=\,\frac{\braket{24}}{\braket{25}\braket{54}}\amp{+-+}{1_{q}\,2^{'}_{\aq}\,4^{'}_g}\,.
\label{eq:D24}
\end{align}
After placing the expressions of shifted anti-holomorphic spinors in  eq.~\eqref{eq:D14} and extracting out the NLP portions we get, 
\begin{align}
\anlp{+-++}{1_{q}\,2_{\aq}\,4_g\,5_g}= & \frac{\braket{14}}{\braket{15}\braket{45}}\left(\frac{2s_{15}}{s_{14}}+\frac{2s_{45}}{s_{14}}
-\frac{\braket{45}[25]}{\braket{14}[12]}\right)\amp{+-+}{1_{q}\,2_{\aq}\,4_g}\,.
\end{align}
This amplitude agree with the sub-leading terms obtained by expanding the first equality in eq.~\eqref{eq:fullqaHgg} around the gluon momentum $p_5$ in the holomorphic limit. Similarly for eq.~\eqref{eq:D24} we get,
\begin{align}
\anlp{+-++}{1_{q}\,2_{\aq}\,5_g\,4_g}= & \frac{\braket{24}}{\braket{25}\braket{54}}\left(\frac{\braket{45}[15]}{\braket{24}[12]}+\frac{2\braket{25}[15]}{\braket{24}[14]}\right)\amp{+-+}{1_{q}\,2_{\aq}\,4_g}\,.
\label{eq:sgamp}
\end{align}
once the exact form of the shifted spinors are inserted. We note in passing that the addition of the two colour-ordered amplitudes given in eqs.~\eqref{eq:D14} and \eqref{eq:D24} can be obtained by using a shifted pair of spinors $|1^{'}]$ and $|2^{'}]$ in the leading order amplitude as,  
\begin{align}
\left.\left(\amp{+-++}{1_{q}\,2_{\aq}\,4_g\,5_g}+\amp{+-++}{1_{q}\,2_{\aq}\,5_g\,4_g}\right)\right|_{\text{LP+NLP}}\,=\,\frac{\braket{12}}{\braket{15}\braket{25}}\amp{+-+}{1^{'}_{q}\,2^{'}_{\aq}\,4_g}\,.
\end{align}
This essentially means that in the next-to-soft gluon limit, the sub-leading colour term of eq.~\eqref{eq:qaqHggsq} has its genesis in the dipole $\mathcal{D}_{12}$ of the leading order amplitude.

\subsection{NLP amplitudes for $ q\aq Q\bar{Q} $}
In this sub-section, we calculate the NLP amplitudes for the scattering process involving Higgs and two pairs of quarks and anti-quarks. In this case, out of the $2^2=4$ helicity configurations, only one is independent. Considering all the particles to be incoming for the process given in eq.~\eqref{eq:qaqHggsq} and the quark with momentum $p_1$ to be soft, the amplitude can be written as, 
\begin{align}
\anlp{+-+-}{1_{q(s)}\,2_{\aq}\,4_Q\,5_{\bar{Q}}}\,=\,\Qqsa{1}{2}\amp{-+-}{2_g\,4_Q\,5_{\bar{Q}}}\,,
\end{align}
where the explicit form of quark operator is provided in Table~\ref{tab:sqtable}. 
In case the  anti-quark with momentum $p_2$ becomes soft, the amplitude takes the following form,
\begin{align}
\anlp{+-+-}{1_{q}\,2_{\aq(s)}\,4_Q\,5_{\bar{Q}}}\,=\,\Qqas{1}{2}\amp{++-}{1_g\,4_Q\,5_{\bar{Q}}}\,,
\end{align}
and as mentioned earlier in sec.~\ref{sec:saq}, $\mathcal{Q}^{q_{h}^{+} \aq_{s}^{-}\to g_{c}^{+}}=\frac{1}{[sh]}$. In the above, we have considered the scenario when flavours of the two $q\bar{q}$ pairs are different. For all (anti-)quarks to be of same flavour, one needs to take into account additional clubbing of soft quark with another adjacent anti-quark and vice versa. Needless to say, one can arrive at the same results when (anti-)holomorphic expansion over the soft (anti-)quark spinor is taken in the full result~\cite{Kauffman:1996ix} and all sub-leading terms are discarded.

\section{NLP logarithms}
\label{sec:nlplog}
To get the NLP logarithms, we need to integrate the squared amplitudes over the unobserved parton phase space in the rest frame of $ p_4 $ and $ p_5 $. We factorise three-body phase space is into two two-body phase spaces in the usual manner: ({\em i}) one containing two coloured particles with momenta $p_4$ and $p_5$, ({\em ii}) another one containing the Higgs ($p_3$) and the joint contribution of the two coloured particles ($p_4+p_5$). We choose the phase space parametrisation in $ d=(4-2\epsilon)$ dimension~\cite{Ravindran:2002dc,Beenakker:1988bq} as, 
\begin{eqnarray}
	p_1&=&(E_1,0,\cdots,0,E_1) \,, \nonumber \\
	p_2&=& (E_2,0,\cdots,0,p_3\sin \psi, p_3\cos \psi - E_1) \,, \nonumber\\
	p_3&=&-(E_3,0,\cdots,0,p_3\sin \psi, p_3\cos \psi ) \,, \nonumber \\
	p_4&=&-\frac{\sqrt{s_{45}}}{2} (1,0,\cdots,0,\sin \theta_1 \sin \theta_2,\sin\theta_1\cos \theta_2,\cos \theta_1) \,, \nonumber\\
	p_5&=&-\frac{\sqrt{s_{45}}}{2} (1,0,\cdots,0,-\sin \theta_1\sin \theta_2,-\sin\theta_1\cos \theta_2,-\cos \theta_1)\,. 
	\label{eq:para1} 
\end{eqnarray}
The helicity dependent differential cross-section at NLP can then be written as, 
\begin{align}
	\left.	s_{12}^2\frac{d^2\sigma^{h_1 h_2 h_3 h_4}}{ds_{13}ds_{23}}\right|_{\text{NLP}}
	\,&= \, \mathcal{F}_{ab} \left (\frac{s_{45}}{\mubar^2}\right )^{-\epsilon}\,\overline{\mathcal{A}_{\text{NLP}}^2} \,,
\label{eq:crossx}	
\end{align}
where
\begin{align}
	\mathcal{F}_{ab}\,=\, \frac{1}{16} \kappa_{ab}\, (N^2-1)\,G^2\left( \frac{\alpha_{s}(\mubar^2)}{4\pi}\right )^2\,
	\left(\frac{s_{13}\,s_{23}-m_H^2\,s_{45}}{\mubar^2\,s_{12}}\right )^{-\E} \,. 
\end{align}
Here $\mubar^2= 4\pi e^ {-\gamma_{E}\epsilon} \mu_r^2$, and $ \kappa_{ab} $ denotes the factor due to the colour average over the initial states {\em i.e.,} $\kappa_{gg}=1/(N^2-1)^{2}$, $\kappa_{q\bar{q}}=1/N^{2}$, $\kappa_{qg}=\kappa_{\bar{q}g}=1/(N(N^2-1))$.
Angular integrated squared amplitude $\overline{\mathcal{A}_{\text{NLP}}^2}$ is defined as,
\begin{align}
	\overline{\mathcal{A}_\text{NLP}^2}\,=&\,\int_{0}^{\pi} d\theta_1\, (\sin\theta_1)^{1-2\E}\int_{0}^{\pi}d\theta_2\, (\sin\theta_2)^{-2\E} \AtsqNLP{} \,.
\end{align}
where $\AtsqNLP{}{}$ is the colour-summed squared NLP amplitude for a given helicity configuration. It is important to take the mass factorisation into account in order for the cancellation of singularities that appear once the above angular integration is performed. 

\subsection{NLP logarithms: $q\aq Hgg$ }
Differential cross-sections would be different for different initial states. Squared NLP colour-ordered amplitudes $\AsqNLP{}{}$ can be obtained through the exchange of momenta applied in the expressions of colour-ordered NLP amplitudes and then squaring them accordingly. For next-to-soft gluon emission $\AsqNLP{}{} = 2 \,\text{Re} (\amp{}{\rm NLP}\amp{\dagger}{\rm LP})$, whereas $\AsqNLP{}{} = (\amp{}{\rm NLP}\amp{\dagger}{\rm NLP})$ for soft (anti-)quark emission. The momentum carried by the (next-to-)soft particle, which can be identified via a suffix $s$ within the bracket, is either $p_4$ or $p_5$, as we choose to work in the $(p_4+p_5)$ rest frame. In the following, we provide the NLP leading logarithms for the independent helicity configurations, as parity conjugate configurations produce same results and the other helicity configurations can be obtained by re-arrenging the external momenta.

\subsubsection{$q\aq$ initiated process}
We define the scattering process as follows, 
\begin{align}
	q(p_{1})+\aq(p_{2})+H(-p_{3})+g(-p_{4})+g(-p_{5})\rightarrow0 & \,\,,
\end{align} 
and consider the gluon with momentum $p_5$ as next-to-soft. As discussed in sec.~\ref{sec:nsg}, there is only one independent helicity configuration that can produce NLP contribution, because the rest are MHV amplitudes and they do not participate at NLP.  

The squared colour-ordered amplitudes that take part in the colour-summed squared amplitude given in eq.~\eqref{eq:qaqHggsq} are,
\begin{align}
&\AsqNLP{+-++}{1_q\,2_{\aq}\,4_g\,5_{g(s)}}\,=\,\left(\frac{3}{s_{15}}+\frac{4}{s_{45}}+\frac{s_{24}}{s_{12} s_{45}}-\frac{s_{14} s_{25}}{s_{12} s_{15} s_{45}}\right)\left|\amp{+-+}{1_q\,2_{\aq}\,4_g}\right|^2\,, \nn
&\AsqNLP{+-++}{1_q\,2_{\aq}\,5_{g(s)}\,4_g}\,=\,\left(-\frac{1}{s_{25}}+\frac{2}{s_{45}}+\frac{s_{14}}{s_{12} s_{45}}-\frac{2 s_{12}}{s_{14} s_{25}}-\frac{s_{15} s_{24}}{s_{25} s_{45} s_{12}}+\frac{2 s_{15} s_{24}}{s_{14} s_{25} s_{45}}\right) \left|\amp{+-+}{1_q\,2_{\aq}\,4_g}\right|^2 \, , \nn
&\left[\amp{+-++}{1_q\,2_{\aq}\,4_g\,5_{g(s)}}+\amp{+-++}{1_q\,2_{\aq}\,5_{g(s)}\,4_g}\right]^2_{\text{NLP}}\,=\, 
\left(-\frac{2}{s_{25}}-\frac{2 s_{24}}{s_{14} s_{25}}+\frac{2 s_{12} s_{45}}{s_{14} s_{15} s_{25}}\right) \left|\amp{+-+}{1_q\,2_{\aq}\,4_g}\right|^2
\, .
\end{align}

Now using eq.~\eqref{eq:crossx}, we calculate the helicity driven NLP leading logarithms that contribute to the differential cross-section {\em i.e.,}
\begin{align}
	\left.	s_{12}^2\frac{d^2\sigma^{+-++}_{1_q\,2_{\aq}\,4_g\,5_{g(s)}}}{ds_{13}\,ds_{23}}\right|_{\text{NLP-LL}}
\,&=\, \mathcal{F}_{q\bar{q}}\,\bigg \{ 4 \pi N \left(\frac{1}{s_{23}}-\frac{1}{s_{13}}\right) \slog \nn
& -\frac{4 \pi }{N} \left [ \left( \frac{1}{s_{23}}-\frac{1}{s_{13}} \right ) \slog+ \frac{2}{s_{23}} \alog \right ] \bigg \} \left|\amp{+-+}{1_q\,2_{\aq}\,4_g}\right|^2. 
\end{align} 
The coefficients of $\slog$ coming from the leading and sub-leading colour terms are exactly same. 

\subsubsection{$gg$ initiated process}
For this process, we assign the momenta of the external particles as follows, 
\begin{align}
	g(p_{1})+g(p_{2})+H(-p_{3})+q(-p_{4})+\aq(-p_{5})\rightarrow0 & \,.
\end{align} 
In this case, either the quark or the anti-quark in the final state can be soft. When the final state quark is soft, the squared colour-ordered amplitudes that are the ingredients of eq.~\eqref{eq:qaqHggsq} can be expressed for the two independent helicity configurations as,
\begin{align}
&\AsqNLP{+++-}{1_g\,2_g\,4_{q}\,5_{\aq(s)}}\,=\, \frac{1}{s_{45}} \left|\amp{+++}{1_g\,2_g\,4_g}\right|^2 
\,, \nn
&\AsqNLP{+++-}{2_g\,1_g\,4_{q}\,5_{\aq(s)}}\,=\, \frac{1}{s_{45}} \left|\amp{+++}{1_g\,2_g\,4_g}\right|^2 
\, , \nn
&\left[\amp{+++-}{1_g\,2_g\,4_{q}\,5_{\aq(s)}}+\amp{+++-}{2_g\,1_g\,4_{q}\,5_{\aq(s)}}\right]^2_{\text{NLP}}\,=\, 0
\, ,
\end{align}
and 
\begin{align}
&\AsqNLP{+-+-}{1_g\,2_g\,4_{q}\,5_{\aq(s)}}\,=\, \frac{1}{s_{45}} \left|\amp{+-+}{1_g\,2_g\,4_g}\right|^2
\,, \nn
&\AsqNLP{+-+-}{2_g\,1_g\,4_{q}\,5_{\aq(s)}}\,=\, \frac{s_{15} s_{24}}{s_{14} s_{25} s_{45}} \left|\amp{+-+}{1_g\,2_g\,4_g}\right|^2
\, , \nn
&\left[\amp{+-+-}{1_g\,2_g\,4_{q}\,5_{\aq(s)}}+\amp{+-+-}{2_g\,1_g\,4_{q}\,5_{\aq(s)}}\right]^2_{\text{NLP}}\,=\, \frac{s_{12}}{s_{14} s_{25}} \left|\amp{+-+}{1_g\,2_g\,4_g}\right|^2
\, .
\end{align}
The NLP leading logarithms that appear in the differential cross-sections of these two independent helicity configurations are given by,
\begin{align}
		\left.	s_{12}^2\frac{d^2\sigma^{+++-}_{1_g\,2_g\,4_{q}\,5_{\aq(s)}}}{ds_{13}\,ds_{23}}\right|_{\text{NLP-LL}}\,&=\,0 \,,
\nn
	\left.	s_{12}^2\frac{d^2\sigma^{+-+-}_{1_g\,2_g\,4_{q}\,5_{\aq(s)}}}{ds_{13}\,ds_{23}}\right|_{\text{NLP-LL}}\,&=\, \mathcal{F}_{gg}\, \Bigg \{ 2 \pi N \left(\frac{s_{12}}{s_{13} s_{23}}\right)\slog -\frac{2 \pi}{N} \left(\frac{s_{12}}{s_{13} s_{23}}\right) \slog \Bigg\} \left|\amp{+-+}{1_g\,2_g\,4_g}\right|^2	\,.
\label{eq:gg2Hqa-qsoft}
\end{align}
Interestingly, no NLP leading logarithm appears for the former helicity configuration and for the latter configuration, no logarithmic term of the form $\alog$ is present in the above equation.  

A similar exercise can be performed when the final state anti-quark is soft. In that case, the squared colour-ordered amplitudes read as, 
\begin{align}
&\AsqNLP{+++-}{1_g\,2_g\,4_{q(s)}\,5_{\aq}}\,=\, \left [
 \left( \frac{1}{s_{24}}+\frac{1}{s_{45}} \right) \left( 1+\frac{s_{12}}{s_{15}}\right)
-\frac{s_{14} s_{25}}{s_{15} s_{24} s_{45}}\right ] \left|\amp{+-+}{2_q\,5_{\aq}\,1_g}\right|^2 
\,, \nn
&\AsqNLP{+++-}{2_g\,1_g\,4_{q(s)}\,5_{\aq}}\,=\, \frac{s_{25}}{s_{15}} \left [
 \left( \frac{1}{s_{14}}+\frac{1}{s_{45}} \right) \left( 1+\frac{s_{12}}{s_{25}}\right)
-\frac{s_{15} s_{24}}{s_{14} s_{25} s_{45}}\right ] \left|\amp{+-+}{2_q\,5_{\aq}\,1_g}\right|^2
\, , \nn
&\left[\amp{+++-}{1_g\,2_g\,4_{q(s)}\,5_{\aq}}+\amp{+++-}{2_g\,1_g\,4_q\,5_{\aq(s)}}\right]^2_{\text{NLP}}\,=\, \left [
 \left( \frac{1}{s_{14}}+\frac{1}{s_{24}} \right) \left( 1+\frac{s_{25}}{s_{15}}\right)
-\frac{s_{12} s_{45}}{s_{14} s_{15} s_{24}}\right ] \left|\amp{+-+}{2_q\,5_{\aq}\,1_g}\right|^2 
\, ,
\end{align}
\begin{align}
&\AsqNLP{+-+-}{1_g\,2_g\,4_{q(s)}\,5_{\aq}}\,= \frac{1}{s_{45} } \left|\amp{+--}{1_g\,2_g\,5_g}\right|^2  
\,, \nn
&\AsqNLP{+-+-}{2_g\,1_g\,4_{q(s)}\,5_{\aq}}\,= \frac{s_{15} s_{24}}{s_{14} s_{25} s_{45}} \left|\amp{+--}{1_g\,2_g\,5_g}\right|^2
\, , \nn
&\left[\amp{+-+-}{1_g\,2_g\,4_{q(s)}\,5_{\aq}}+\amp{+-+-}{2_g\,1_g\,4_q\,5_{\aq(s)}}\right]^2_{\text{NLP}}\,=\, 
\frac{s_{12}}{s_{14} s_{25}}
\left|\amp{+--}{1_g\,2_g\,5_g}\right|^2
\, ,
\end{align}
and the NLP leading logarithms of the differential cross-sections can be expressed as,
\begin{align}
	\left.	s_{12}^2\frac{d^2\sigma^{+++-}_{1_g\,2_g\,4_{q(s)}\,5_{\aq}}}{ds_{13}\,ds_{23}}\right|_{\text{NLP-LL}}\,&=\, \mathcal{F}_{gg}\, \Bigg \{ 2 \pi N \left(\frac{1}{s_{13}}+\frac {s_{13}}{s_{23}^2}\right)\slog \nn &-\frac{2 \pi}{N}\left[\left( \frac{1}{s_{13}}+\frac {s_{13}}{s_{23}^2}\right) \slog -\frac{2 }{s_{23}} \alog \right]  \Bigg\} \left|\amp{+-+}{2_q\,5_{\aq}\,1_g}\right|^2 \,, \nn
	\left.	s_{12}^2\frac{d^2\sigma^{+-+-}_{1_g\,2_g\,4_{q(s)}\,5_{\aq}}}{ds_{13}\,ds_{23}}\right|_{\text{NLP-LL}}\,&=\, \mathcal{F}_{gg}\, \Bigg \{ 2 \pi N \left(\frac{s_{12}}{s_{13} s_{23}}\right)\slog -\frac{2 \pi}{N} \left(\frac{s_{12}}{s_{13} s_{23}}\right) \slog \Bigg\}	\left|\amp{+--}{1_g\,2_g\,5_g}\right|^2 \,.
\end{align}
 Unlike eq.~\eqref{eq:gg2Hqa-qsoft}, the first helicity configuration in the above produces a non-zero result when the final state anti-quark is soft, whereas the result for the latter configuration matches to the soft quark case except for the overall leading order squared amplitude.   

\subsubsection{$qg$ initiated process}
Here the process we are interested in is described as,  
\begin{align}
	g(p_{1})+q(p_{2})+H(-p_{3})+\aq(-p_{4})+g(-p_{5})\rightarrow0 & \,.
\end{align}
Note that either of the final state gluon or quark can be (next-to-)soft and they can contribute to the NLP threshold cross-section. We first consider the case when the gluon is next-to-soft and as discussed earlier, no MHV configuration would play any role here. There is only one independent helicity configuration and to get the colour-summed squared amplitude of that configuration, we calculate the following colour-ordered squared amplitudes,
\begin{align}
&\AsqNLP{++-+}{1_g\,2_q\,4_{\aq}\,5_{g(s)}}\,=\,\AsqNLP{+-++}{1_q\,2_{\aq}\,5_{g(s)}\,4_g} \Big\{ 1\to2, 2\to4, 4\to1 \Big\}
\,, \nn
&\AsqNLP{++-+}{5_g(s)\,2_q\,4_{\aq}\,1_{g}}\,=\,\AsqNLP{+-++}{1_q\,2_{\aq}\,4_{g}\,5_{g(s)}} \Big\{ 1\to2, 2\to4, 4\to1 \Big\}
\, , \nn
&\left[\amp{++-+}{1_g\,2_q\,4_{\aq}\,5_{g(s)}}+\amp{++-+}{5_g(s)\,2_q\,4_{\aq}\,1_{g}}\right]^2_{\text{NLP}}\,=\, 
\left[\amp{+-++}{1_q\,2_{\aq}\,5_{g(s)}\,4_{g}}+\amp{+-++}{1_q\,2_{\aq}\,4_{g}\,5_{g(s)}}\right]^2_{\text{NLP}} \Big\{ 1\to2, 2\to4, 4\to1 \Big\}
\, .
\end{align}
The colour-summed squared amplitude thus obtained takes part in the computation of the differential cross-section and the NLP leading logarithms of that cross-section can be written as, 
\begin{align}
\xsctn{++-+}{1_g\,2_q\,4_{\aq}\,5_{g(s)}}\,=&\,\mathcal{F}_{qg}\, \bigg\{4 \pi N\left[\left(\frac{1}{s_{13}}+\frac{4}{s_{23}}\right)\slog-\frac{1}{s_{13}}\alog\right] \nn &-\frac{4 \pi }{N}\left(\frac{1}{s_{13}}\right) \slog\bigg\} \left|\amp{+-+}{2_q\,4_{\aq}\,1_g}\right|^2 \,.
\end{align}

Now we provide the results when the final state quark is soft. Following are the expressions of the squared colour-ordered amplitudes for the two independent helicity configurations,
\begin{align}
&\AsqNLP{++-+}{1_g\,2_q\,4_{\aq(s)}\,5_{g}}\,=\, \AsqNLP{+++-}{1_g\,2_g\,4_{q}\,5_{\aq(s)}} \Big\{ 1\to5, 2\to1, 4\to2, 5\to4 \Big\}
\,, \nn
&\AsqNLP{++-+}{1_g\,2_q\,4_{\aq(s)}\,5_{g}}\,=\,  \AsqNLP{+++-}{2_g\,1_g\,4_{q}\,5_{\aq(s)}} \Big\{ 1\to5, 2\to1, 4\to2, 5\to4 \Big\}
\, , \nn
&\left[\amp{++-+}{1_g\,2_q\,4_{\aq(s)}\,5_{g}}+\amp{++-+}{1_g\,2_q\,4_{\aq(s)}\,5_{g}}\right]^2_{\text{NLP}}\,=\, 
\left[\amp{+++-}{1_g\,2_g\,4_{q}\,5_{\aq(s)}}+\amp{+++-}{2_g\,1_g\,4_{q}\,5_{\aq(s)}}\right]^2_{\text{NLP}}\ \Big\{ 1\to5, 2\to1, 4\to2, 5\to4 \Big\}
\, ,
\end{align}
and 
\begin{align}
&\AsqNLP{-+-+}{1_g\,2_q\,4_{\aq(s)}\,5_{g}}\,=  \AsqNLP{+-+-}{1_g\,2_g\,4_{q}\,5_{\aq(s)}} \Big\{ 1\to5, 2\to1, 4\to2, 5\to4 \Big\}
\,, \nn
&\AsqNLP{-+-+}{5_g\,2_q\,4_{\aq(s)}\,1_{g}}\,= \AsqNLP{+-+-}{2_g\,1_g\,4_{q}\,5_{\aq(s)}} \Big\{ 1\to5, 2\to1, 4\to2, 5\to4 \Big\}
\, , \nn
&\left[\amp{-+-+}{1_g\,2_q\,4_{\aq(s)}\,5_{g}}+\amp{-+-+}{5_g\,2_q\,4_{\aq(s)}\,1_{g}}\right]^2_{\text{NLP}}\,=\, 
\left[\amp{+-+-}{1_g\,2_g\,4_{q}\,5_{\aq(s)}}+\amp{+-+-}{2_g\,1_g\,4_{q}\,5_{\aq(s)}}\right]^2_{\text{NLP}}\ \Big\{ 1\to5, 2\to1, 4\to2, 5\to4 \Big\}
\, .
\end{align}

The NLP leading logarithms are then given by,
\begin{align}
	\xsctn{++-+}{1_g\,2_q\,4_{\aq(s)}\,5_{g}}\,&=\, \mathcal{F}_{qg}\, \bigg\{4 \pi N\left(\frac{1}{s_{13}}\right)\slog\bigg\} \left|\amp{+++}{1_g\,2_g\,5_g}\right|^2 \,, \nn
	\xsctn{-+-+}{1_g\,2_q\,4_{\aq(s)}\,5_{g}}\,&=\, \mathcal{F}_{qg}\, \bigg\{2 \pi N\left[\left(\frac{3}{s_{13}}\right)\slog+\frac{2}{s_{13}}\alog\right] \nn 
	&-\frac{2 \pi }{N}\left(\frac{1}{s_{13}}\right) \slog\bigg\} \left|\amp{-++}{1_g\,2_g\,5_g}\right|^2 \,,
\label{eq:qgHqg-qsoft}
\end{align}
where the sub-leading colour term does not contribute to the \lq$+\!+\!-\!\;+$\rq\, configuration. 

\subsubsection{$\aq g$ initiated process}
To describe this process we consider the following assignment of external momenta,  
\begin{align}
	\aq(p_{1})+g(p_{2})+H(-p_{3})+g(-p_{4})+q(-p_{5})\rightarrow0 & \,.
\end{align}
As described in the previous sub-section, next-to-soft gluon emission leads to only one independent helicity configuration, whereas soft anti-quark in the final state causes two independent helicity structures to appear at the NLP.  The results for the squared colour-ordered amplitudes and the NLP leading logarithms are first provided for the next-to-soft gluon and then for the soft quark emission. 

For the next-to-soft gluon emission, 
\begin{align}
&\AsqNLP{-+++}{1_{\aq}\,2_g\,4_{g(s)}\,5_{q}}\,=\,\AsqNLP{+-++}{1_q\,2_{\aq}\,4_g\,5_{g(s)}} \Big\{ 1\to5, 2\to1, 4\to2, 5\to4 \Big\}
\,, \nn
&\AsqNLP{-+++}{1_{\aq}\,4_{g(s)}\,2_g\,5_{q}}\,=\,\AsqNLP{+-++}{1_q\,2_{\aq}\,5_{g(s)\,4_g}} \Big\{ 1\to5, 2\to1, 4\to2, 5\to4 \Big\}\, , \nn
&\left[\amp{-+++}{1_{\aq}\,2_g\,4_{g(s)}\,5_{q}}+\amp{-+++}{1_{\aq}\,4_{g(s)}\,2_g\,5_{q}}\right]^2_{\text{NLP}}\,=\, 
\left[\amp{+-++}{1_{q}\,2_{\aq}\,4_{g}\,5_{g(s)}}+\amp{+-++}{1_{q}\,2_{\aq}\,5_{g(s)}\,4_{g}}\right]^2_{\text{NLP}}
\Big\{ 1\to5, 2\to1, 4\to2, 5\to4 \Big\}\
\, ,
\end{align}
and,
\begin{align}
\xsctn{-+++}{1_{\aq}\,2_g\,4_{g(s)}\,5_{q}}\,=&\,\mathcal{F}_{\bar{q}g}\, \bigg\{4 \pi N\left[\left(\frac{4}{s_{13}}-\frac{1}{s_{23}}\right)\slog+\left(\frac{2}{s_{13}}-\frac{1}{s_{23}}\right)\alog\right] \nn &+\frac{4 \pi}{N}\left(\frac{1 }{s_{23}}\right) \slog\bigg\} 
 \left|\amp{+-+}{5_q\,1_{\aq}\,2_g}\right|^2\,.
\end{align}
Such results for the final state soft quark read as,
\begin{align}
&\AsqNLP{-+++}{1_{\aq}\,2_g\,4_g\,5_{q(s)}}\,=\, \AsqNLP{+++-}{1_g\,2_g\,4_{q(s)}\,5_{\aq}} \Big\{ 1\to2, 2\to4, 4\to5, 5\to1 \Big\}
\,, \nn
&\AsqNLP{-+++}{1_{\aq}\,4_g\,2_g\,5_{q(s)}}\,=\,  \AsqNLP{+++-}{2_g\,1_g\,4_{q(s)}\,5_{\aq}} \Big\{ 1\to2, 2\to4, 4\to5, 5\to1 \Big\}
\, , \nn
&\left[\amp{-+++}{1_{\aq}\,2_g\,4_g\,5_{q(s)}}+\amp{-+++}{1_{\aq}\,4_g\,2_g\,5_{q(s)}}\right]^2_{\text{NLP}}\,=\, 
\left[\amp{+++-}{1_g\,2_g\,4_{q(s)}\,5_{\aq}}+\amp{+++-}{2_g\,1_g\,4_{q(s)}\,5_{\aq}}\right]^2_{\text{NLP}}  \Big\{ 1\to2, 2\to4, 4\to5, 5\to1 \Big\}
\, ,
\end{align}
\begin{align}
&\AsqNLP{-+-+}{1_{\aq}\,2_g\,4_g\,5_{q(s)}}\,=  \AsqNLP{+-+-}{1_g\,2_g\,4_{q(s)}\,5_{\aq}} \Big\{ 1\to2, 2\to4, 4\to5, 5\to1 \Big\}
\,, \nn
&\AsqNLP{-+-+}{1_{\aq}\,4_g\,2_g\,5_{q(s)}}\,= \AsqNLP{+-+-}{2_g\,1_g\,4_{q(s)}\,5_{\aq}} \Big\{ 1\to2, 2\to4, 4\to5, 5\to1 \Big\}
\, , \nn
&\left[\amp{-+-+}{1_{\aq}\,2_g\,4_g\,5_{q(s)}}+\amp{-+-+}{1_{\aq}\,4_g\,2_g\,5_{q(s)}}\right]^2_{\text{NLP}}\,=\, 
\left[\amp{+-+-}{1_g\,2_g\,4_{q(s)}\,5_{\aq}}+\amp{+-+-}{2_g\,1_g\,4_{q(s)}\,5_{\aq}}\right]^2_{\text{NLP}}  \Big\{ 1\to2, 2\to4, 4\to5, 5\to1 \Big\}
\, ,
\end{align}
and,
\begin{align}
	\xsctn{-+++}{1_{\aq}\,2_g\,4_g\,5_{q(s)}}\,&=\, \mathcal{F}_{\bar{q}g}\, \bigg\{2 \pi N\left[\left(\frac{2 s_{13}}{s_{12} s_{23}}+\frac{s_{23}}{s_{12} s_{13}}\right)\slog-\left(\frac{2}{s_{12}}\right)\alog\right] \nn 
	&-\frac{2 \pi}{N}\left(\frac{s_{23}}{s_{12} s_{13}}\right) \slog\bigg\} 
 \left|\amp{+-+}{4_q\,1_{\aq}\,2_g}\right|^2 \,, \nn
	\xsctn{-+-+}{1_{\aq}\,2_g\,4_g\,5_{q(s)}}\,&=\, \mathcal{F}_{\bar{q}g}\, \bigg\{2 \pi N\left[\left(\frac{3}{s_{23}}\right)\slog+\frac{2}{s_{23}}\alog\right] \nn 
	&-\frac{2 \pi }{N}\left(\frac{1}{s_{23}}\right) \slog\bigg\} \left|\amp{-+-}{1_g\,2_g\,4_g}\right|^2 \,.
\end{align}
As compared to eq.~\eqref{eq:qgHqg-qsoft}, the results for the \lq$-\!+\!-\!\;+$\rq\, configuration are related to each other by $s_{13}\leftrightarrow s_{23}$ interchange due to their inherent association at the leading order in the soft limit, whereas this is not the case for the other helicity configuration.  

\subsection{NLP logarithms: $q\aq HQ\bar{Q}$ }
The $q\aq$ initiated process that produces one Higgs boson and another pair of quark and anti-quark can be written as,
\begin{align}
q(p_1) + \aq (p_2) + H(-p_3)+Q (-p_4) +\bar{Q} (-p_5) \rightarrow 0\,,
\end{align}
where the flavour of the final state pair can be same or different as compared to the initial state $q\aq$ pair. Here either of the final state quark or anti-quark can be soft, thereby leading to NLP leading logarithms at the differential cross-section level. There is only one independent helicity configuration and the expression of the colour-summed squared amplitude is given in eq.~\eqref{eq:flavoursq}. It is evident that all the leading and sub-leading colour terms exist only when the quarks are of same flavour. Considering the final state quark to be soft, the colour-ordered squared amplitudes are as follows,  
\begin{align}
&\AsqNLP{+-+-}{1_q\,2_{\aq}\,4_{Q}\,5_{\bar{Q}(s)}}\,=\,\frac{1}{s_{45}} \left|\amp{+-+}{1_q\,2_{\aq}\,4_g}\right|^2 \,, \nn
&\AsqNLP{+-+-}{1_q\,2_{\aq}\,4_{q}\,5_{{\aq}(s)}}\,=\,\frac{s_{12}}{s_{15} s_{24}} \left|\amp{+-+}{1_q\,2_{\aq}\,4_g}\right|^2 \,, \nn
&\left[\amp{+-+-}{1_q\,2_{\aq}\,4_{Q}\,5_{{\bar{Q}}(s)}}\amp{\dagger\,+-+-}{1_q\,2_{\aq}\,4_{q}\,5_{{\aq}(s)}}
+ c.c. \right]_{\text{NLP}}
\,=\,  \left(\frac{1}{s_{45}} + \frac{s_{12}}{s_{15} s_{24}} - \frac{s_{14} s_{25}}{s_{15} s_{24} s_{45}} \right) \left|\amp{+-+}{1_q\,2_{\aq}\,4_g}\right|^2
\,.
\end{align} 
The NLP leading logarithm results are given by,
\begin{align}
\xsctn{+-+-}{1_q\,2_{\aq}\,4_{q}\,5_{{\aq}(s)}}\,=\, \mathcal{F}_{q\bar{q}}\, \bigg\{ 2 \pi N \bigg(\frac{s_{12}}{s_{13} s_{23}}\bigg) \slog
\bigg\} \left|\amp{+-+}{1_q\,2_{\aq}\,4_g}\right|^2 \,.
\end{align}
Similar calculations for the final state anti-quark to be soft result into, 
\begin{align}
&\AsqNLP{+-+-}{1_q\,2_{\aq}\,4_{Q(s)}\,5_{\bar{Q}}}\,=\,\frac{1}{s_{45}} \left|\amp{+--}{1_q\,2_{\aq}\,5_g}\right|^2 \,, \nn
&\AsqNLP{+-+-}{1_q\,2_{\aq}\,4_{q(s)}\,5_{\aq}}\,=\,\frac{s_{12}}{s_{15} s_{24}} \left|\amp{+--}{1_q\,2_{\aq}\,5_g}\right|^2 \,, \nn
&\left[\amp{+-+-}{1_q\,2_{\aq}\,4_{Q(s)}\,5_{{\bar{Q}}}}\amp{\dagger\,+-+-}{1_q\,2_{\aq}\,4_{q(s)}\,5_{{\aq}}}
+ c.c. \right]_{\text{NLP}}
\,=\, \left(\frac{1}{s_{45}} + \frac{s_{12}}{s_{15} s_{24}} - \frac{s_{14} s_{25}}{s_{15} s_{24} s_{45}} \right) \left|\amp{+--}{1_q\,2_{\aq}\,5_g}\right|^2
\,.
\end{align} 
and,
\begin{align}
\xsctn{+-+-}{1_q\,2_{\aq}\,4_{q(s)}\,5_{\aq}}\,=\, \mathcal{F}_{q\bar{q}}\, \bigg\{ 2 \pi N \bigg(\frac{s_{12}}{s_{13} s_{23}}\bigg) \slog 
\bigg\} \left|\amp{+--}{1_q\,2_{\aq}\,5_g}\right|^2 \,.
\end{align}
Note that, except the squared leading order amplitudes, all expressions both in the soft quark and anti-quark cases are same. In the NLP leading logarithms, no contribution comes from the sub-leading colour term and the non-zero leading colour term exists only when all the quarks and anti-quarks are of same flavour.

\section{Psuedo-scalar Higgs plus jet production}
To explore the universality of the NLP logarithms obtained in the previous section, the natural and straightforward extension is to study the case when the Higgs boson is replaced by a pseudo-scalar Higgs. In the large top mass limit when the top quark effect is integrated out, the effective Lagrangian for this case can be written as, 
\begin{align}
	\mathcal{L}_{\text{eff}}\,=\, G_a \,A\, \text{Tr} (\tilde{F}^a_{\mu\nu} F^{\mu\nu, a})\,
\end{align}
where $A$ represents the pseudo-scalar Higgs, the coupling $G_a =\alpha_{s}/(8\pi v)$ and $\tilde{F}^a_{\mu\nu}=\frac{1}{2}\epsilon^{\mu\nu\lambda\rho} {F}^a_{\lambda\rho}$. As the pseudo-scalar does not carry any colour charge, colour orderings of the helicity amplitudes remain exactly same to the scalar Higgs plus jet production process given in section~\ref{sec:corder}. 

Pseudo-scalar Higgs plus four gluon amplitudes are identical to that of Higgs plus four gluon amplitudes for all non-NMHV helicity configurations, however the NMHV amplitudes differ due to a relative sign change in one of the two terms~\cite{Kauffman:1999ie}. Following~\cite{Pal:2023vec}, it is already known that Higgs plus four gluon NMHV amplitudes do not contribute at NLP leading logarithmic order and we find the same holds true here as well for the pseudo-scalar Higgs case. As a consequence, forms of all NLP leading logarithms for the pseudo-scalar Higgs plus jet production match to the Higgs plus jet production process for each and every helicity configuration~\cite{Pal:2023vec}.   

Expressions of the scattering amplitudes that include a $q\bar{q}$ pair and two gluons along with a Higgs or pseudo-scalar Higgs are same when the gluons posses same helicities. When the two gluons carry opposite helicities, they differ by a relative sign change between the two terms in the last two identities given in eq.~\eqref{eq:fullqaHgg}. Recall that NLP leading logarithms for next-to-soft gluon radiation can only arise when the gluons have same helicities, as MHV amplitudes do not play any role here. This feature ensures that the structures of NLP leading logarithms for the psuedo-scalar Higgs plus jet production are identical to that of Higgs plus jet production when the radiated gluon is next-to-soft. For soft quark or anti-quark radiation, the forms of NLP leading logarithms for the Higgs and pseudo-scalar Higgs cases appear to be same because -- ($i$) the first identity of eq.~\eqref{eq:fullqaHgg} remains same in both the cases, ($ii$) any one of the two terms in the last two identities of eq.~\eqref{eq:fullqaHgg} contributes at NLP, thereby diminishing the effect of relative sign changes in them. 

For the process involving a Higgs or pseudo-scalar Higgs in association with two $q\bar{q}$ pairs, helicity amplitudes again differ by a relative sign change in one of the two terms as given in ~\cite{Kauffman:1999ie}. However this change in sign does not alter the forms of NLP leading logarithms, as one of the two types of terms takes part in the NLP calculation at a time depending on whether the quark or anti-quark is soft. 

Through this study we establish the universality of NLP leading logarithms for the case of Higgs and pseudo-scalar Higgs production in association with a final state jet. Numerical estimations of these logarithms depend on the values of coupling $G_a$ and mass of the pseudo-scalar Higgs $m_A$, which can consistently be absorbed in $\mathcal{F}_{ab}$ of eq.~\eqref{eq:crossx}. When masses of Higgs and pseudo-scalar Higgs are considered to be equal, one can readily scale the NLP results for the Higgs plus one jet production by a factor of $(G_a^2/G^2)$ and obtain them for the pseudo-scalar Higgs plus one jet production.

\section{Conclusions}
\label{sec:conclu}
Enormous amount of data from the LHC demands precise predictions from perturbative QCD. The required precision is achievable by computing higher order corrections and performing resummation originating from the singular region of the phase space. NLP resummation is one of the many different ways to achieve percent-level accuracy in the theoretical predictions. To build a consistent resummation method for NLP logarithms, one needs to calculate many process at this accuracy to test the validity of universality at this order. The draught of NLP results for jet-associated processes in the literature indicates the need for a better method to calculate them.

In this paper, we have focused on computing the NLP leading logarithms for all possible channels involving (anti-)quarks for the Higgs plus one jet production in a hadron collider. These NLP logarithms originate when either a next-to-soft gluon or a soft (anti-)quark is emitted from the leading-order process. To obtain the results, we have utilised a method involving shifted spinors, as outlined in~\cite{Pal:2023vec}, to calculate the next-to-soft gluon corrections. Additionally, we have introduced for the first time soft (anti-)quark operators based on spinor variables to account for the NLP effects arising from soft (anti-)quark radiation in Higgs plus one jet production. Our analysis has revealed that MHV amplitudes for processes involving a $q\bar{q}$ pair alongside a Higgs and two gluons do not contribute at the NLP in the next-to-soft gluon limit. Together with the findings in~\cite{Pal:2023vec}, our results provide NLP leading logarithms for all the channels that take part in Higgs plus one jet production. The absence of MHV amplitudes for the above mentioned channel and NMHV amplitudes for Higgs plus four-gluons channel at NLP indicates the need for further exploration, particularly from the generic structural point of view in scattering amplitudes.

The universality of NLP logarithms and their resummation is an active area of research and many different approaches have been proposed in the literature to explore that direction. One of the ways to understand this universality is to calculate many processes with increasing complexities and investigate the structure of NLP logarithms. Although colour singlet production processes exhibit a universal pattern, the universality of NLP terms for processes with coloured final states is not yet in place. In this article, we have observed the universal nature of NLP leading logarithms when production of a scalar or a pseudo-scalar Higgs is considered in association with a final state jet. We believe that the simplicity of the shifted dipole spinors technique of~\cite{Pal:2023vec} along with the soft (anti-)quark spinor operator technique developed in this paper will speed up the calculation of NLP logarithms for many more jet-associated processes relevant at the LHC.

\section*{Acknowledgements}
We thank V. Ravindran for useful discussions and gratefully acknowledge the hospitality of IMSc during the preparation of this manuscript. We also thank R. K. Ellis for carefully reading the manuscript and providing useful suggestions. 
SS is supported in part by the SERB-MATRICS under Grant No. MTR/2022/000135.
 \bibliographystyle{bibstyle}
\bibliography{ref}

\end{document}